\documentstyle[12pt]{article}
\newcommand{\bce}{\begin{center}}
\newcommand{\ece}{\end{center}}
\newcommand{\beq}{\begin{equation}}
\newcommand{\eeq}{\end{equation}}
\newcommand{\bea}{\vspace{0.25cm}\begin{eqnarray}}
\newcommand{\eea}{\end{eqnarray}}

\newcommand{\ba}{\begin{array}}
\newcommand{\ea}{\end{array}}

\newcommand{\integ}{\int\!d}

\newcommand{\ket}[1]{| {#1} \rangle}
\newcommand{\bra}[1]{\langle {#1} |}
\newcommand{\ave}[1]{\langle {#1} \rangle}

\newcommand{\doublespace}{
    \renewcommand{\baselinestretch}{1.6}\large\normalsize}
\def\lsim{\mathrel{\rlap{\lower4pt\hbox{\hskip1pt$\sim$}}
    \raise1pt\hbox{$<$}}}     %less than or approx. symbol
\def\gsim{\mathrel{\rlap{\lower4pt\hbox{\hskip1pt$\sim$}}
    \raise1pt\hbox{$>$}}}     %greater than or approx. symbol

\begin{document}
\vspace{1.0in}
\begin{flushright}
{\small Dec, 1993}
\end{flushright}
\vspace{2.0cm}
\begin{center}
{\Large{\bf Chaos Driven Decay of Nuclear Giant Resonances: \newline
Route to Quantum Self-Organization}}
\vspace{1.0cm}

S. Dro\.zd\.z$^{a,b,c}$,
S. Nishizaki$^{c,d}$,
and J. Wambach$^{a,c}$
\vspace{1.0cm}

{\it
 a) Department of Physics, University of Illinois at Urbana,
IL 61801, USA\newline
 b) Institute of Nuclear Physics, PL - 31-342 Krak\'ow, Poland \newline
 c) Institut f\"ur Kernphysik, Forschungszentrum J\"ulich,
D-5170 J\"ulich, Germany \newline
 d) College of Humanities
and Social Sciences, Iwate University, Ueda  3-18-34,\newline
Morioka 020, Japan}
\end{center}
\vskip 1cm
\abstract
{The influence of background states with increasing level of complexity
on the strength distribution of the isoscalar and isovector giant quadrupole
resonance in $^{40}$Ca is studied. It is found that
the background characteristics, typical for chaotic systems, strongly
affects the fluctuation properties of the strength distribution.
In particular, the small components of the wave function obey
a scaling law analogous to self-organized systems at the critical state.
This appears to be consistent with the Porter-Thomas distribution of the
transition strength.}
\vspace{1.0cm}

\newpage
\doublespace

Nuclear physics has contributed significantly to the recent progress
in understanding the chaotic aspects of nonlinear dynamical systems,
especially in the context of classical versus quantum correspondence.
The nucleus is particularly well suited for such studies
due to the intrinsic quantum nature of the nuclear
many-body problem on the one hand, and the wealth of experimental data on
the other. The chaotic nature of the
nucleus is by now well documented empirically \cite{BFF}
and seems natural,  bearing in mind the many-body
character of the nucleus and the complicated form of the nucleon-nucleon
interaction. Still, however,  nuclei reveal many
regular, collective phenomena and this coexistence of chaos and
collectivity is a challenging problem for quantitative study \cite{GW}.
Especially interesting in this respect are the nuclear giant resonances
which carry a large fraction of the total transition strength and
are located many MeV above the ground state, in the energy region which
is expected to be dominated by chaotic dynamics.

The giant resonance, as a short time phenomenon, involves simple
configurations of one-particle one-hole (1p-1h) type.
Chaos may influence the subsequent decay of these components which
occurs on longer time scales and gradually evolves to more and more
complex configurations. Eventually, the initial energy deposited in the
nucleus is redistributed over
all available degrees of freedom and the limit of the compound nucleus is
reached. This is the limit of fully developed chaos and, as a result,
quantum stochastic methods based on the random matrix theory
\cite{VWZ} or, alternatively, molecular dynamical approaches generating
chaotic behavior \cite{SODB}, prove appropriate.
The process of giant resonance formation and its subsequent
decay towards the compound nucleus occurs in a closed system and the most
basic approach is in terms of a single Hamiltonian acting
in a rich enough Hilbert space such that the relevant degrees of freedom are
included. This also provides the most natural scheme for the coupling
between a collective state and the complex background. In quantum
mechanical terms one can then speak about the large and the small
components of the nuclear wave function \cite{Sol}.
It is the purpose of the present paper to work out such a scheme,
to identify which ingredients are relevant, to study the role of chaos
on the giant resonance physical observables and finally,
from a more general perspective, to contribute
to the understanding of the universal aspects of the coexistence between
collectivity  and chaos in small many-body quantum systems.

We start from the nuclear Hamiltonian which, in second quantized form, reads
as
\beq
\hat H=\sum_i\epsilon_i a_i^{\dag} a_i+{1\over 2}\sum_{ij,kl}v_{ij,kl}
a_i^{\dag} a_j^{\dag} a_la_k.
\eeq
The first term denotes the mean field which represents its most regular
part \cite{Zel} while the second term is the residual interaction. Clearly
a diagonalization of $\hat H$ in the full Hilbert space of n-particle
n-hole excitations is numerically prohibitive, and may not be necessary.
Our recent study \cite{DNSW} of level statistics in the subspace
generated by two-particle two-hole (2p-2h) excitations has shown that
it is sufficient to diagonalize the Hamiltonian within a truncated
subspace of 1p-1h and 2p-2h states \cite{DNS}:
\beq
\ket{1}\equiv a_p^{\dag} a_h\ket{0};\quad \ket{2}\equiv
a_{p_1}^{\dag} a_{p_2}^{\dag}  a_{h_2}a_{h_1}\ket{0}
\eeq
For the present discussion it is more transparent to prediagonalize $v$ in
the 1p-1h and 2p-2h subspaces such that
\beq
\ket{\tilde 1}=\sum_1C^{\tilde 1}_1\ket{1};\quad \ket{\tilde 2}=\sum_2
C^{\tilde 2}_2\ket{2}
\eeq
Then the Schr\"odinger equation takes the following form
\beq
\left ( \begin{array}{cc}
E_{\tilde 1}&A_{\tilde 1\tilde 2}\\
A^*_{\tilde 2 \tilde 1}&E_{\tilde 2}\end{array}\right )
\left ( \begin{array}{c}
X^N_{\tilde 1}\\X^N_{\tilde 2}\end{array}\right )=E_N
\left ( \begin{array}{c}
X^N_{\tilde 1}\\X^N_{\tilde 2}\end{array}\right )
\label{eq:Schr}
\eeq
where $E_{\tilde 1}$ and $E_{\tilde 2}$ denote the energy eigenvalues
in the 1p-1h and 2p-2h subspaces respectively and $A_{\tilde 1\tilde 2}=
\sum_{1 2}C^{\tilde 1}_1\bra{1}v\ket{2}C^{\tilde 2}_2$ mediates the coupling
between these two spaces. The solutions of Eq.~(\ref{eq:Schr}) yield the
transition strength distribution $S_F(E)$ in response to an external field
$\hat F=\sum_{ij}F_{ij}a^{\dag}_ia_j$ as
\beq
S_F(E)=\sum_{\tilde 1,N}|X^N _{\tilde 1}\bra{0}\hat F\ket{\tilde 1}|^2
\delta(E-E_N)
\eeq
with its energy moments being defined as
\beq
\ave{E^n}=\integ ES_F(E)E^n/ \integ ES_F(E).
\eeq

For the specific case of the quadrupole response in $^{40}$Ca,
considered here, we have chosen the mean field and residual interaction
as in ref.~\cite{SW} including two major shells above and below the Fermi
level. Our study is based on an explicit diagonalization in Eq.~(4) which
involves more than 11,000 states out of which only 26 are 1p-1h states.
Computational restrictions require to limit the number of 2p-2h states. It
turns out that, including those up to 50 MeV excitation energy,
suffices for a realistic description of the measured response function
\cite{Wou}. This yields altogether 3014  $2^+$ states which is numerically
manageable. The results displayed in Fig.~1 and Fig.~2 yield a
mean excitation energy $\ave{E}$ of 30.84 MeV for the isovector and 24.45
MeV for the isoscalar transitions, independent of the mixing with 2p-2h
states. The latter can be easily understood
by writing the mean energy as $\ave{E}=\bra{0}\hat F^{\dagger}[\hat H,\hat F]
\ket{0}$ and realizing that $\ket{0}$ is given by a Slaterdeterminant
of occupied single-particle states.
For the energy dispersion, $\sigma=(\ave{E^2}-\ave{E}^2)^{1/2}$,
we obtain 3.41 MeV and 4.28 MeV when only 1p-1h excitations are considered.
Allowing for the coupling to 2p2h-states these values are increased to
5.40 MeV and 5.42 MeV, respectively, independent of whether effects
of the residual interaction in the 2p-2h subspace is included or not.
The latter can be deduced from the form of $\ket{0}$ and the fact
that
$\ave{E^2}=-\bra{0}[\hat H,\hat F^{\dagger}][\hat H,\hat F]\ket{0}$.
Motivated by our previous results \cite{DNSW} on the level fluctuations
in the prediagonalized 2p-2h space we distinguish three cases:
(1) no residual interaction in this space (the corresponding strength
distribution is shown in Figs.~1b and 2b) for which the nearest-neighbor
spacing distribution is sharply peaked near zero, because of degeneracies,
(2) the inclusion of particle-particle and hole-hole two-body matrixelements
(Figs.~1c and 2c) which results in a Poissonian distribution,
characteristic of the
known universality class of generic integrable systems \cite{BT} (3)
the use of the full residual interaction (Figs.~1d and 2d) which yields
the fluctuations of the Gaussian orthogonal ensemble (GOE) \cite{BFF}
characteristic of classically chaotic systems. All three cases introduce
significant modifications of the 1p-1h 'doorway' strength distribution
(notice the change in magnitude of the
transition matrixelements) resulting in a gradual reduction of the
large components accompanied by a simultaneous amplification of the
smaller ones when going from case (1) to case (3). The isovector response
is affected more than the isoscalar one especially when going from (2)
to (3) (Figs.~1c and 1d and Figs.~2c and 2d).

Of course, what is important for the degree of mixing between the spaces
spanned by $\ket{\tilde 1}$ and $\ket{\tilde 2}$ is not
only the spectral properties of the eigenenergies in $\ket{\tilde 2}$ but
also the distribution of the coupling matrix elements
$A_{{\tilde 1}{\tilde 2}}$. This distribution can be influenced by the
degree of coherence in the wave packet $\ket{F}=\hat F\ket{0}$,
initially formed by an external field. The motion in the isoscalar case
is more coherent than in the isovector case since
protons and neutrons move in phase. Therefore, the isoscalar state is
expected to be more resistive against decay than its isovector counterpart
\cite{DNS}. This tendency is also seen from Fig.~3 which displays
the number of matrix elements $A_{F{\tilde 2}}$ of given
magnitude, connecting the state $\ket F$ and the states $\ket{\tilde 2}$.
In the cases referred to as (1) and (2) above, additional selection rules
lead to large degeneracies and consequently the corresponding distributions
sharply peak at zero. When the full residual interaction
is used in generating the vectors $\ket{\tilde 2}$
(case (3)) these selection rules are removed and there are no
vanishing coupling matrix elements any more. In the isovector case
the wings of distribution seem to be consistent with a Gaussian as one
would expect for a random process. In the isoscalar case, however,
$A_{F{\tilde 2}}$ remains more localized around
small values and a Gaussian fit to the wings is unsatisfactory.

The analysis of the distribution of the mixing matrix elements
is consistent with the results seen in Figs.~1 and 2;
the isovector strength distribution is more affected by the
complex background. Another and perhaps the most interesting effect is
that the isovector strength is distributed much more uniformly, not only in
excitation energy but also in magnitude.
A bit of imagination may even suggest a certain kind of
self-similarity regarding the clustering and the relative size of the
transitions. The picture in Fig.~1d becomes
reminiscent of a self-organized system at its
critical state \cite{BTW,KNWZ} where the equilibrium balance reduces the
dimensionality. This observation finds confirmation in more quantitative
terms. Fig.~4 shows the total number $N$ of transitions of
magnitude smaller than a given threshold value $S_{th}$, as a function
of $S_{th}$. For the isovector resonance in the chaotic case (Fig.~1d)
we find, except for the largest transitions, a scaling law of the form
$N \sim S_{th}^{\alpha}$ ($\alpha \approx 0.50$) (indicated by the
straight line fit in the upper part of Fig.~4) which indeed signals a
reduction of dimensionality.  The two nonchaotic cases (Figs.~1b and c)
display
a more complicated behavior. It is interesting to notice that a similar
scaling applies also to the isoscalar resonance (lower part of Fig.~4)
even though this is not so obvious from Fig.~2d. This time, however, the
scaling interval is somewhat shorter and $\alpha \approx 0.48$.

The strength distribution can be considered as an
attractor for the decay process starting out of equilibrium.
It is nothing but the Fourier transform of the time correlation function
$\ave {F(0)| F(t)}$ which, in the form of an envelope \cite{Hel},
describes the process of gradual convergence to such an attractor and,
thus, resolves it.
This sets the parallel to a procedure \cite{DOS}
which resolves the self-similarity and the scale invariance
in classical chaos. This analogy provides further arguments
for interpreting the above scaling law as another manifestation of
'1/f'-type behavior \cite{DH}. So far, such a behavior has been
identified mostly
on the classical level \cite{All} for a variety of observables and models.
In the present case of an 'avalanche' of the decaying giant resonance
it is reflecting the chaotic properties of a strictly
quantum mechanical phenomenon.

The influence of chaos on the strength distribution is believed
\cite{AL,MDVB} to manifest itself in a Porter-Thomas distribution \cite{PT}
of transition strengths maximizing the entropy of the strength
distribution \cite{AL}. Converting the appropriate (differential)
Porter-Thomas distribution to a cumulative allows a direct comparison
with our calculated distributions. The results are given by the thin
solid lines in Fig.~3. They show the same scaling behavior with
$\alpha=1/2$ exactly as can be easily seen from the analytical
form of the Porter-Thomas distribution.
One thus finds an impressive consistence. The strength distributions
in Figs.~1 and 2 represent the coexistence of collectivity and chaos
and, therefore, the amplified abundance of large components as compared
to a purely random process is natural.
Being more coherent, for the isoscalar
resonance the  onset of the scaling regime is delayed. It is also
interesting to see that even a fully random process,
as represented by the Porter-Thomas
distribution, develops a trace of such a delay. The possibility of a
similar two-phase behavior has been identified \cite{BZ} in random
cascade models.

\vskip 1.5cm
\begin{center}
{\bf Acknowledgement}
\end{center}

This work was supported in part by the Polish KBN grants 2 2409 9102
and 2 P302 157 04 and by NSF grant PHY-89-21025.
One of the authors (S. N.) would like to express his thanks to
the Alexander von Humboldt Foundation for a fellowship.

\newpage
\vspace{.25in}
\parindent=.0cm            %align refs

\newpage
\begin{center}
{\bf Figure Captions}
\end{center}

\begin{itemize}
\item[{\bf Fig.~1}] The isovector quadrupole strength distribution in
$^{40}$Ca: (a) no coupling to the 2p-2h subspace (b) no residual
interaction in 2p-2h subspace (c) including only particle-particle and
hole-hole matrixelements in the diagonalization of the 2p-2h subspace
(d) diagonalization of the full residual interaction in the 2p-2h
subspace.

\item[{\bf Fig.~2}] Same as Fig.~1 but for the isoscalar quadrupole
strength distribution.

\item[{\bf Fig.~3}] The distribution of coupling matrixelements of the
isovector and isoscalar
wave packets $\ket{F}$ to 2p-2h excitations: (a) with no
residual interaction in the 2p-2h subspace, (b) including only pp- and hh
matrixelements in this space, (c) diagonalizing the residual interaction
fully. The full lines denote a Gaussian fit to these distributions.

\item[{\bf Fig.~4}] The total number $N$ of transitions of given
strength below a threshold value $S_{th}$ as a function of $S_{th}$.
The open triangles refer to the case (b), the open squares to the case (c)
and thick dots to the case (d) of Figs.~1 and 2, respectively.
The solid line indicates the best fit to the later case. The thin solid
lines represents the same quantity determined from a Porter-Thomas
distribution.

\end{itemize}

\end{document}